\documentclass[fleqn,usenatbib]{mnras}

\usepackage{newtxtext,newtxmath}

\usepackage[T1]{fontenc}
\usepackage{ae,aecompl}

\usepackage{graphicx}	
\usepackage{amsmath}	
\usepackage{amssymb}	
\usepackage{color}    
\usepackage{soul}

\defcitealias{sell15b}{S15}

\title[Chandra X-ray Constraints on SN~2016hnk]{Chandra X-ray Constraints on the Candidate Ca-rich Gap Transient SN~2016hnk}

\author[P. H. Sell et al.]{
P. H. Sell,$^{1,2}$\thanks{E-mail: psell@physics.uoc.gr (PHS)}
K. Arur,$^{3}$
T. J. Maccarone,$^{3}$
R. Kotak,$^{4}$
C. Knigge,$^{5}$
D. J. Sand$^{6}$
\newauthor \ and S. Valenti$^{7}$
\\
$^{1}$Foundation for Research and Technology-Hellas, 71110 Heraklion, Crete, Greece\\
$^{2}$Physics Department \& Institute of Theoretical \& Computational Physics,University of Crete, 71003 Heraklion, Crete, Greece\\
$^{3}$Department of Physics, Texas Tech University, Lubbock, TX 79409, USA\\
$^{4}$Astrophysics Research Centre, School of Mathematics \& Physics, Queen's University Belfast, BT7 1NN, UK\\
$^{5}$School of Physics and Astronomy, University of Southampton, Highfield, Southampton, SO17 1BJ, UK\\
$^{6}$Department of Astronomy/Steward Observatory, 933 North Cherry Avenue, Room N204, Tucson, AZ 85721-0065, USA\\
$^{7}$Department of Physics, University of California, 1 Shields Avenue, Davis, CA 95616-5270, USA
}

\date{Accepted XXX. Received YYY; in original form ZZZ}

\pubyear{2018}

\begin{document}
\label{firstpage}
\pagerange{\pageref{firstpage}--\pageref{lastpage}}
\maketitle

\begin{abstract}

We present a Chandra observation of SN~2016hnk, a candidate Ca-rich gap transient.  This observation was specifically designed to test whether or not this transient was the result of the tidal detonation of a white dwarf by an intermediate-mass black hole.  Since we detect no X-ray emission 28 days after the discovery of the transient, as predicted from fall-back accretion, we rule out this model.  Our upper limit of $\sim 10$~M$_\odot$ does not allow us to rule out a neutron star or stellar-mass black hole detonator due limits on the sensitivity of Chandra to soft X-rays and unconstrained variables tied to the structure of super-Eddington accretion disks.  Together with other Chandra and multiwavelength observations, our analysis strongly argues against the intermediate-mass black hole tidal detonation scenario for Ca-rich gap transients more generally.

\end{abstract}




\section{Introduction}
\label{section:introduction}

In approximately the last decade as time-domain astronomy has radically expanded, a wealth of discoveries of new types of transient sources have been revealed.  A fraction of them fall in the gap between novae and supernovae \citep[$-10 \lesssim M_V \lesssim -16$~mag, e.g.,][]{kasliwal11}.  One of these new types of sources is the class of Ca-rich gap transients, so called because they have peak luminosities in the gap between novae and supernovae ($-15 \lesssim M_R \lesssim -16$) and show unusually strong optical calcium emission a few weeks after the peak luminosity \citep{kasliwal12}.

As the list of these known sources continues to grow \citep{sullivan11,kasliwal12,valenti14,lunnan17}, a set of common characteristics has been established for this class of objects.  They additionally lack hydrogen lines similar to Type~I supernovae, have characteristic velocities of 6000~--~11000~km~s$^{-1}$, evolve slightly faster than Type~I supernovae (rise times $\leq 15$~days), and appear to be offset from star-forming regions and the stellar light of their host galaxies \citep{yuan13}.  Deep HST imaging indicates that they do not seem to be associated with star clusters or dwarf galaxies, but may have been kicked \citep{lyman13,lyman14,lyman16}, a finding seemingly consistent with large line-of sight velocity shifts seen in their nebular spectra \citep[$100 \lesssim v_{los} \lesssim 2000$~km~s$^{-1}$,][]{foley15}.

The physical nature of the progenitor is far from resolved, and a variety of models have been invoked to explain this phenomenon.  Most focus on the nucleosynthetic products observed as the transient ages, primarily the lack of iron-peak elements and the strong calcium emission observed at the start of the nebular phase.

Perhaps the most common interpretation involves the interaction with one or two white dwarfs \citep[WDs;][]{garcia-berro17}, as was suggested by \citet{perets10} for SN~2005E.  Many simulations \citep{woosley11,waldman11,dan15,meng15,dessart15} have shown that intermediate elements and peak luminosities are produced by the detonation of an accreted He shell reaching a critical mass or when two low-mass WDs collide.  The runaway accretion leading to a detonation can occur, for instance, in AM~CVn systems \citep{bildsten07,shen09,shen10}.  To produce the correct nucleosynthetic yields, modeling usually focuses on the partial or total destruction of a low-mass helium WD, where the total mass fused will be smaller than that for common types of Type Ia supernovae.  The discovery of increasing numbers of low-mass helium WDs in the field \citep{marsh95,kepler07} and globular clusters \citep{taylor01,strickler09} make this a viable scenario, even as the search to find the lowest-mass WDs that must form through binary interactions continues \citep{liebert04,kilic07,vennes11,hermes13}.

The source of the detonation or fusion could also involve a more compact object, a black hole (BH) or neutron star (NS).  Compression by extreme tidal forces could fuse the helium in the detonation of the WD \citep{rosswog09} or in a nuclear-dominated accretion flow \citep[NuDAF;][]{metzger12,fernandez13,margalit16}.  The encounter could could be induced via a random flyby of an intermediate-mass BH (IMBH) in a dense stellar cluster or a three-body interaction in a triple system \citep[see][for an extensive discussion; hereafter S15]{sell15b}.  Then a fraction of the shredded WD material will fall back onto the NS or BH, producing copious UV/X-ray emission in the accretion flow.

However, \citet{milisavljevic17} has recently challenged the low-mass helium WD paradigm and the homogeneity of Ca-rich gap transients with multiwavelength observations of a particular Ca-rich transient, iPTF15eqv.  Their analysis shows that a stripped envelope stellar progenitor star seems to explain their observations better than a WD.  Such a model involving a star of initial mass $M \sim 10$~M$_\odot$ was also used to explain SN~2005cz (though not clear that this was a Ca-rich gap transient), which has intriguing implications for common envelope evolution \citep{kawabata10,suh11}.  This would indicate the strong presence of calcium in the nebular spectrum is not caused by a single type of explosion mechanism.

Our focus here is on testing the model of the tidal detonation of a low-mass WD by a NS or intermediate-/stellar-mass BH by detecting X-rays from the fall-back accretion.  This was first done with a follow-up {\em Chandra} X-ray observation of SN~2012hn \citepalias{sell15b}, but the very late time of the observation led to upper limits that did not place strong constraints on the model.  We test this model again with an X-ray observation of SN~2016hnk, which is designed to be much more sensitive (see Section~\ref{section:observations}).

Based on extensive photometric and spectral monitoring and on initial classification criteria, SN~2016hnk qualifies as a candidate Ca-rich gap transient \citep[Lluis Galbany priv. comm.; ATels: \#9685, \#9703, \#9704, and \#9705,][]{tonry16,cannizzaro16,dimitriadis16,pan16}\footnote{https://wis-tns.weizmann.ac.il/object/2016hnk}$^{,}$\footnote{http://www.astronomerstelegram.org/}:   no hydrogen, $6000 < v < 11000$~km~s$^{-1}$, a discovery absolute magnitude of $-16.1$, very strong calcium lines in the nebular spectrum.  In general, it is a peculiar Type Ia supernova described as being most similar to PTF09dav.

This similarity complicates the classification of the source, as PTF09dav is somewhat of an outlier in the Ca-rich gap transient class.  As has been remarked before \citep{valenti14,foley15,meng15,white15}, PTF09dav spectroscopically resembled 1991bg-like SNe Ia, had lower velocity ejecta ($v \sim 6000$~km~s$^{-1}$) than other Ca-rich gap transients ($v \sim 11000-12000$~km~s$^{-1}$) and exhibited scandium, strontium, and hydrogen \citep[the latter from previous nova explosions?;][]{sullivan11,kasliwal12}.  \citet{dessart15} postulated that this could stem from variations in burning yields.  A detailed analysis of SN~2016hnk will be undertaken in a future publication (Galbany et al. in prep.).

We describe our observations in Section~\ref{section:observations}.  Then we describe our analysis of the physical and spectral models used in Section~\ref{section:analysis}.  Finally, we conclude with a brief discussion of the implications of this work in Section~\ref{section:conclusions}.

\section{Observations}
\label{section:observations}

As soon as we became aware that SN~2016nhk was a likely Ca-rich gap transient (approximately 1 week after its discovery), we activated our {\em Chandra} trigger of the source.  Keeping in mind that the tidal detonation model predicts $L \propto t^{-5/3}$, our observations were planned with a balance between sensitivity and feasibility.  Given the predictions in \citet{rosswog09} and \citetalias{sell15b}, we should be able to detect such an event as long as $M_{BH} \gtrsim 10$~M$_\odot$ and $D \lesssim 100$~Mpc (see Section~\ref{section:analysis}).  A ``medium" trigger (15--30~days response time) was therefore the slowest we could tolerate given a reasonable 30~ksec exposure time.

We observed SN~2016hnk with the {\em Chandra X-ray Observatory} at 11:34:36\,UT on 2016-11-24 (OBS\_ID=18022), 28.01 days after the detection of the source (11:22:33\,UT on 2016-10-27).  We assume the first observation is at the peak in the subsequent analysis.  As long as the time of the explosion is within a few weeks of the detection (a very reasonable assumption given peak luminosities of other Ca-rich gap transients and their decay rates), the upper limits would change no more than a factor of three as shown in Figure~\ref{fig:upper_limits} and our conclusions will not change.

Using CIAO~4.8 and CALDB 4.7.2 \citep{fruscione06}, we reprocessed the level=1 event file with the {\sc chandra\_repro} tool with the default settings to apply the most recent calibrations to create a level=2 event file, resulting an exposure time of 26667~s.  No source was detected at the position of the transient:  RA,Dec=33.319292$^\circ$,-7.661306$^\circ$; we checked that the $1 \sigma$ coordinate alignment uncertainty between the optical and X-ray images is $\leq 0.4^{\prime \prime}$.  The transient is coincident with the galaxy, MCG-01-06-070, which is at a distance of $\sim 70$~Mpc assuming a standard LCDM cosmology given its redshift, $z=0.0160$.  It is projected approximately $12.6^{\prime \prime}$ or 4.1~kpc from the nucleus of the galaxy, near the edge of the bulge.

To report upper limits on source fluxes and luminosities for our following analysis, we extract with the {\sc specextract} tool the source and local background spectra at the reported position of the source.  We use a circle of radius $2^{\prime \prime}$ for the source spectrum.  For the background spectrum, we use a radius of $10^{\prime \prime}$, masking the region of the source spectrum and a nearby bright source $3.8^{\prime \prime}$ from the source position.  We first calculate the source count rate upper limit using the area-weighted scaling of the background counts.  Taking into account the proper Poisson uncertainties following \citet{kraft91} and using the exposure time above, we find an upper limit of $1.99 \times 10^{-4}$ counts per second (0.5--8.0~keV at 99.73\% confidence).  We describe our detailed spectral analysis from which we calculate upper limits to the luminosity in the next section.

\section{Analysis}
\label{section:analysis}

\begin{figure}
\includegraphics[width=\columnwidth,trim=2.5cm 0cm 2cm 0cm]{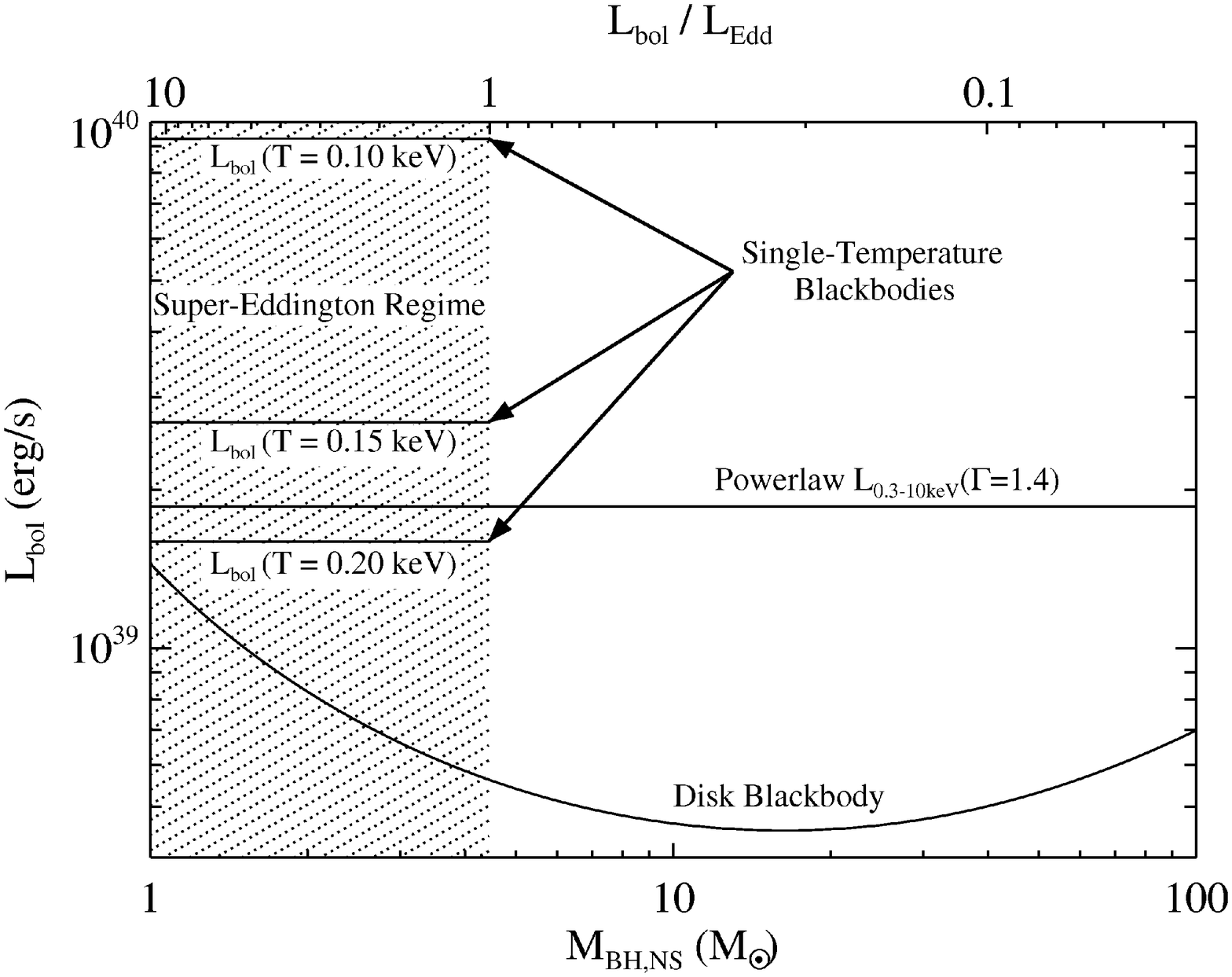}
\includegraphics[width=\columnwidth,trim=2.5cm 2cm 2cm 1cm]{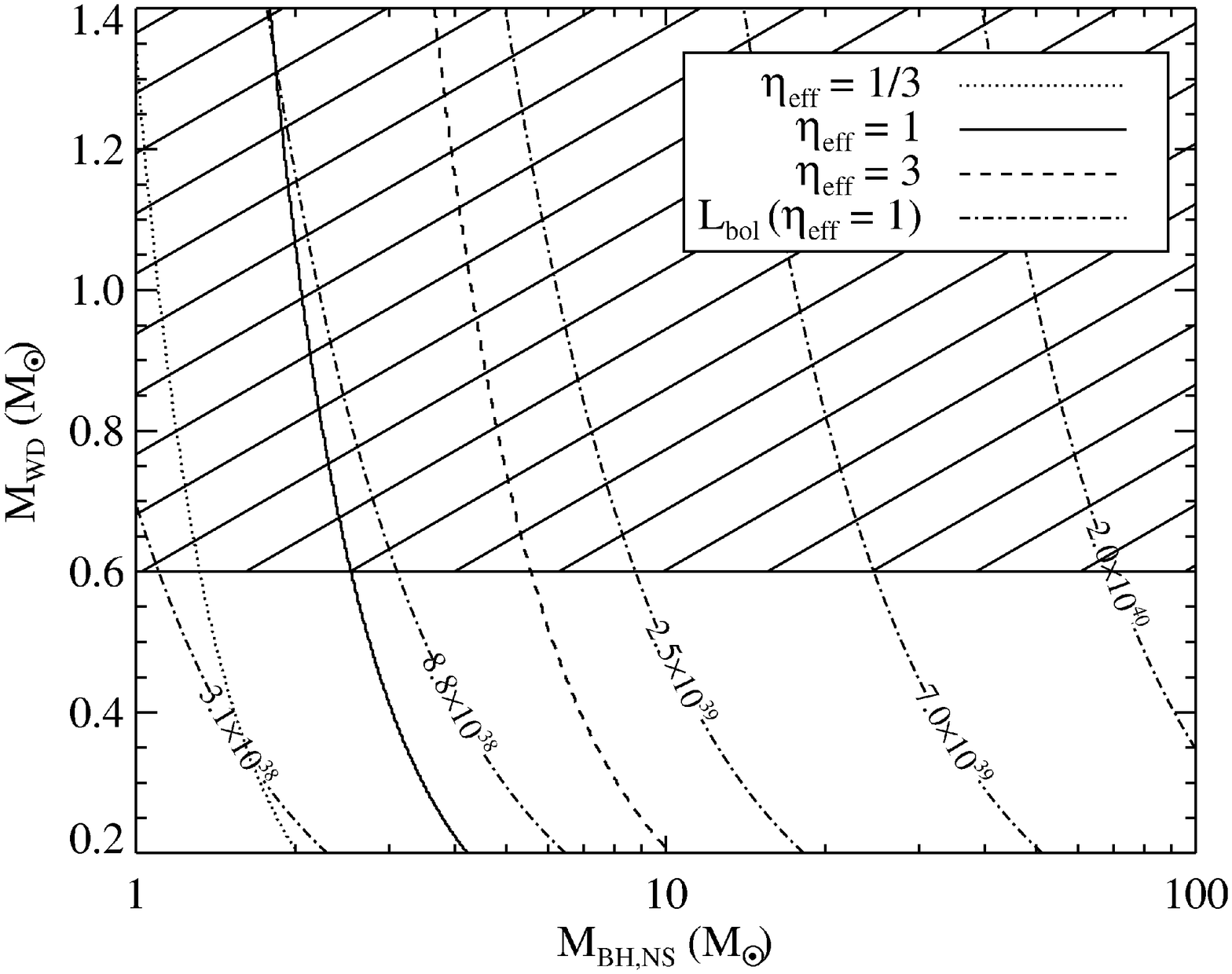}
\caption{{\bf Top}:  The solid curved line is the upper limit to the bolometric luminosity as a function of black hole mass and bolometric luminosity for the disk blackbody model.  We also show upper limits assuming a couple other simple models:  single temperature blackbodies with temperatures typical of sources observed in the super-Eddington state and a powerlaw with a photon index typical for accretion flows.  This also emphasizes the strongly varying luminosity limit at low temperatures.   {\bf Bottom:}  We compare the disk blackbody $L_{bol}$ upper limits to luminosities in the fallback accretion model to construct upper limits to $M_{BH,NS}$ and $M_{WD}$ spanning a range of a factor of 9 in effective efficiency ($\eta_{eff}$).  We overplot the bolometric luminosity predictions from the fallback accretion model for reference (dot-dashed lines in erg~s$^{-1}$).  $M_{WD}>0.6$ is not allowed for this model because it produces too many heavy elements.}
\label{fig:upper_limits}
\end{figure}

Our primary goal is to test the tidal detonation model for this candidate Ca-rich gap transient.  We do this by comparing the model predicted luminosity to the luminosity upper limits calculated from the X-ray data.  We essentially repeated the analysis from \citetalias{sell15b} but with a more careful consideration of the bolometric correction ($L_\textrm{bol} / L_\textrm{0.5~--~8.0~keV}$) using a few of the simplest, common spectral models for accretion disks.

First, we briefly summarize our model construction and assumptions.  When a WD passes too close to a NS or BH (not supermassive or else it is swallowed whole), tidal forces compress, shred, and, to produce the observed intermediate-mass elements, detonate the WD material.  After the initial encounter, a fraction of the WD material falls back on the more compact object.  We calculate the fallback accretion rate and efficiency from \citet{li02} for the case of a WD and a NS or BH (instead of a star by a supermassive black hole as originally formulated).  We then make the same simplifying assumptions as in \citetalias{sell15b} for the mass-radius relation of a WD, the fraction of the WD mass that falls back to the BH or NS \citep[f=0.35 as suggested by][]{rosswog09}, and the standard accretion disk state transitions, which we summarize in the next paragraph.

However, because we observed the source almost exactly 4 weeks after the transient instead of approximately almost 1.5 years as in \citetalias{sell15b}, we are orders of magnitude more sensitive to any X-ray emission associated with fall-back accretion.  The upper limit we are able to place on the mass of any compact BH/NS detonator is also correspondingly smaller.  Given the large range of BH/NS masses that we are now sensitive to ($M\approx 1$~--~$2 \times 10^{5}~$M$_\odot$, the physically allowed limits for BHs/NSs for a tidal disruption), the bolometric correction changes strongly with the structure of the accretion disk:  the optically thin and geometrically thick disk of an advection-dominated accretion flow \citep[ADAF, $L\leq0.02$L$_\textrm{Edd}$,][]{esin97}; the optically thick and geometrically thin disk \citep[$0.02$L$_{Edd}\le L \le$L$_\textrm{Edd}$,][]{shakura73}; and the optically thick and geometrically thick disk in the super-Eddington state as typically seen for ultra-luminous X-ray sources \citep[ULXs; $L>$L$_\textrm{Edd}$, e.g.,][]{feng11,narayan17}.

Given the fallback accretion model and the timescale at which we observed the transient, the upper limits on $M_{BH,NS}$ require that $L_{bol} / L_{Edd} \sim 1$ within factors of order unity.  In this case, the disk will be in the so-called high-soft or very high (super-Eddington) state.  For the high-soft state, we use a disk blackbody model \citep[{\sc xsdiskbb}, e.g.,][]{makishima00}.  For the very high state, we use a single temperature blackbody \citep[{\sc xsbbody}, e.g.,][]{urquhart16}.  With each model, we include absorption from the Galactic foreground \citep[{\sc phabs},][]{dickey90}.  We fit the spectrum from 0.5-8~keV using {\sc wstat} in Sherpa \citep{freeman01}.  For each model, we calculate joint $3\sigma$ upper limits on the model parameters and then bolometric luminosities from these limits.  In Fig.~\ref{fig:upper_limits}, we show the upper limits that bracket reasonable ranges in spectral hardness seen in accretion disks.  We also show the constraints that would result if the X-ray emission was described as a conservatively hard powerlaw of $\Gamma = 1.4$ for jet/coronal emission over the energy range that {\em Chandra} is well-calibrated \citep[{\sc xspowerlaw},][]{roberts16}.  Different disk inclinations reveal different spectral features, especially in the super-Eddington state.

If we consider the most extreme temperatures observed for ULXs, $T \sim 0.05$~keV, then the bolometric correction is huge ($\sim 100$).  However, our upper limits allow only mild super-Eddington accretion, $\dot{M}_{Edd} \lesssim 10$, not the most extreme Eddington accretion rates suggested for some ULXs \citep[$\dot{M}_{Edd} \sim 300$,][]{feng16}.  Given these super-Eddington models, this makes our choice of $T=0.1$~keV quite conservative.  Even in this extreme case, the bolometric correction is only a factor of 4 at most.  In any case, even allowing for such large bolometric corrections and uncertainties in the shape of the disk spectrum, we can set a firm upper limit of $M_{detonator} \lesssim 10$~M$_\odot$ $\times$ ($L_{bol} / L_{Edd}$).  This still strongly rules out an IMBH as the detonator.

Though we have constrained the range of reasonable bolometric correction factors, there are many other parameter uncertainties that we cannot constrain from the data.  We have subsumed these uncertainties in the efficiency parameter, $\eta_{eff}$, which encodes the fall-back accretion fraction, the BH spin, the inclination angle of the disk ($L_{bol} \propto \cos(i)$), the spectral hardening and radial correction factors for a disk blackbody in \citet{makishima00}, accretion efficiency parameters (radiative, advective, disk wind and jet), choice of spectral model, other details on the formulation of the model, uncertainties in the distance to the source, etc.  As in \citetalias{sell15b} and as shown in Figure~\ref{fig:upper_limits}, we allow for a factor of 3 uncertainty in $\eta_{eff}$.  Given the magnitude by which we rule out the IMBH tidal detonation scenario, constructing a combination of these parameters that allows this model given the observations would have to be extremely contrived.

\section{Discussion and Conclusions}
\label{section:conclusions}

Using the upper limit to the X-ray luminosity from a {\em Chandra} observation of SN~2016hnk, we have tested and ruled out the model for the tidal detonation of a low-mass WD by an IMBH for this recent candidate Ca-rich gap transient.  We first compare our results to {\em Chandra} observations of two other Ca-rich transients.

In \citetalias{sell15b}, a follow-up {\em Chandra} observation of the Ca-rich gap transient, SN~2012hn, did not detect the fall-back accretion expected for this model.  However, given the $L \propto t^{-5/3}$ and the late time at which we observed the transient (533~days), the observed upper limits were not nearly as constraining.  \citet{milisavljevic17} carried out a similar analysis with their follow-up {\em Chandra} observations of iPTF15eqv.  Since this transient was also not detected in X-rays, we simply repeat the analysis from section~\ref{section:analysis} and find an upper limit similar to that derived for our source SN~2016hnk ($M_{BH} \sim 20$~M$_\odot$), consistent with their analysis within factors of order unity.  Though their {\em Chandra} observation was approximately a factor of three shorter (9.9~ksec) and the delay time was approximately a factor of 2 longer ($\sim 83$~days, assuming a light curve evolution similar to other bona-fide Ca-rich gap transients), the $M_{BH}$ upper limit is similar because the object was much closer (30~Mpc).

These three follow-up {\em Chandra} observations as well as the lack of (dense) star clusters at the positions of these objects \citep[e.g.,][]{lyman16} together all strongly argue against the IMBH tidal detonation model for Ca-rich gap transients.  However, because of the large uncertainties in the structure of the accretion flow at super-Eddington mass accretion rates (see section~\ref{section:analysis}) and many possible ways three-body interactions could drive a low-mass WD too close to a NS or stellar-mass BH \citepalias{sell15b}, there is no way to conclusively rule out the low-mass ($M_{BH,NS} \lesssim 10$~M$_\odot$) detonator scenario from the current data.  In addition, this is a computationally and physically difficult problem to simulate, and there is a very large, uncertain range of model parameter space to explore \citepalias[e.g.,][]{sell15b}.  Exactly under what conditions a low-mass helium WD in these simulations will partially or fully detonate is still an open question \citep{holcomb13,vick16,tanikawa17}.

Furthermore, other proposed scenarios listed in Section~\ref{section:introduction} cannot be ruled out based on the lack of X-ray emission.  For a NuDAF caused by an encounter by a WD and NS, though the amount of calcium produced is too low, the disk structure is extremely uncertain \citep{margalit16}.  Nevertheless, simple predictions based on the the mass accretion rate ($L \simeq 0.1 \dot{M} c^2$) and expected decay curve suggest that the predicted X-ray luminosity would be many orders of magnitude below our sensitivity limits.  The lack of X-ray emission for the other single or double progenitor scenarios is also not surprising, given that $L_X \lesssim 10^{38}$~erg~s$^{-1}$ for models of Type~Ia supernovae \citep[e.g.,][]{dmitriadis14}.

If the progenitor system for Ca-rich gap transients is not the result of a tidal detonation by a NS or BH, then have we yet seen this case in other transient sources?  Other multiwavelength observations of strange X-ray transients and supernovae provide some evidence that we might have \citep{krolik11,shcherbakov13,jonker13}.  There is considerable interest in constructing detailed simulations of this phenomenon accounting for line \citep{clausen11} to broadband emission, resulting in comparable predicted X-ray luminosities as our model in section~\ref{section:analysis} \citep[or larger in the case of seeing a source of beamed emission;][]{haas12,macleod16}.  These new simulations together with tighter constraints from the next generation of X-ray satellites should yield further insights in the near future.

\section*{Acknowledgements}

We acknowledge the important work of the various transient factories and Astronomer's Telegrams that reported the detection and initial details of this source.  This work has been funded through {\em Chandra} grant GO6-17066X.  Research by DJS is supported by NSF grants AST-1412504 and AST-1517649.




\bibliographystyle{mnras}
\bibliography{ms} 


\bsp	
\label{lastpage}
\end{document}